\documentclass[]{aa}
\usepackage{txfonts}
\usepackage{graphicx}
\usepackage{natbib}

\begin{document}

\title{Physical parameters and $\pm 0.2$\% parallax of the detached eclipsing binary
V923 Scorpii\thanks{Based on observations made with ESO telescopes at the Paranal Observatory, 
under ESO program 091.D-0207 and data from the MOST satellite, a former Canadian Space Agency mission,
jointly operated by Microsatellite Systems Canada Inc. (MSCI; formerly Dynacon
Inc.), the University of Toronto Institute for Aerospace Studies, and the University
of British Columbia; assistance was provided by the University of Vienna.}}
\titlerunning{Spectro-photo-interferometric analysis of V923~Sco}
\authorrunning{T. Pribulla et al.}
\author{
T.~Pribulla\inst{1}\thanks{ESO visiting scientist}
\and
A.~M\'erand\inst{2}
\and
P.~Kervella\inst{3,4}
\and
C.~Cameron\inst{5,6}
\and
C.~Deen\inst{7,8}
\and
P.J.V.~Garcia\inst{9}
\and
M.~Horrobin\inst{10}
\and
J.M.~Matthews\inst{11}
\and
A.F.J.~Moffat\inst{12}
\and
O.~Pfuhl\inst{8}
\and
S.M.~Rucinski\inst{13}
\and
O.~Straub\inst{4}
\and
W.W.~Weiss\inst{14}
}
\offprints{T. Pribulla}
\mail{pribulla@ta3.sk}
\institute{
Astronomical Institute, Slovak Academy of Sciences, 059 60 Tatransk\'a Lomnica, Slovakia
\and
European Southern Observatory, Alonso de C\'ordova 3107, Casilla 19001, Santiago 19, Chile
\and
Unidad Mixta Internacional Franco-Chilena de Astronom\'{i}a (CNRS UMI 3386),
Departamento de Astronom\'{i}a, Universidad de Chile, Camino El Observatorio 1515,
Las Condes, Santiago, Chile
\and
LESIA (UMR 8109), Observatoire de Paris, PSL Research University, CNRS, UPMC, Univ.
Paris-Diderot, 5 Place Jules Janssen, 92195 Meudon, France
\and
Department of Mathematics, Physics and Geology, Cape Breton University, 1250
Grand Lake Road, Sydney, Nova Scotia, Canada, B1P 6L2
\and
Canadian Coast Guard College, Department of Arts, Sciences, and
Languages, Sydney, Nova Scotia, B1R 2J6, Canada
\and
Max-Planck-Institut f\"ur Astronomie, K\"onigstuhl 17, 69117 Heidelberg, Germany
\and
Max Planck Institute for extraterrestrial Physics, Giessenbachstr., 85748 Garching, Germany
\and
Faculdade de Engenharia, Universidade do Porto, rua Dr. Roberto Frias, 4200-465 Porto, Portugal;
CENTRA -- Centro de Astrof\'{\i}sica e Gravita\c c\~{a}o, IST, Universidade de Lisboa, P-1049-001 Lisboa, Portugal
\and
Physikalisches Institut, Universit\"at zu K\"oln, Z\"ulpicher Str. 77, 50937 K\"oln, Germany
\and
Department of Physics and Astronomy, University of British Columbia, Vancouver, BC V6T1Z1, Canada
\and
D\'epartment de physique, Universit\'e de Montr\'eal C.P. 6128,
Succursale Centre-Ville, Montr\'eal, QC H3C 3J7, Canada
\and
Department of Astronomy and Astrophysics, University of Toronto, 50 St George Street, Toronto, ON M5S 3H4, Canada
\and
University of Vienna, Institute for Astronomy, T\"urkenschanzstrasse 17, A-1180 Vienna, Austria}

\date{Received ; Accepted}
\abstract
{V923~Sco is a bright ($V$ = 5.91), nearby ($\pi$ = 15.46$\pm$0.40 mas) southern eclipsing binary. Because both components are slow rotators, the minimum masses of the components are known with 0.2\% precision from spectroscopy. The system seems ideal for very precise mass, radius, and luminosity determinations and, owing to its proximity and long orbital period ($\sim$ 34.8 days), promises to be resolved with long-baseline interferometry.}
{The principal aim is very accurate determinations of absolute stellar parameters for both components of the eclipsing binary and a model-independent determination of the distance.}
{New high-precision photometry of both eclipses of V923~Sco with the MOST satellite was obtained. The system was spatially resolved with the VLTI AMBER, PIONIER, and GRAVITY instruments at nine epochs. Combining the projected size of the spectroscopic orbit (in km) and visual orbit (in mas) the distance to the system is derived. Simultaneous analysis of photometric, spectroscopic, and interferometric data was performed to obtain a robust determination of the absolute parameters.}
{Very precise absolute parameters of the components were derived in spite of the parameter correlations. The primary component is found 
 to be overluminous for its mass. Combining spectroscopic and interferometric observations enabled us to determine the distance to V923~Sco with better than 0.2 \% precision, which provides a stringent test of Gaia parallaxes.}
{It is shown that combining spectroscopic and interferometric observations of nearby eclipsing binaries can lead to extremely accurate parallaxes and stellar parameters.}


\maketitle

\section{Introduction}

HR6327 ($V$ = 5.91) was found to be a SB2 system by \citet{bennet67}. The authors determined an orbital period of $P$ = 34.8189 days and estimated the spectral types of its components as F3IV-V and F3V. Using HR6327 as a comparison star, \citet{bolton76} serendipitously discovered a primary eclipse in the system of depth 0.35 mag. \citet{kholopov81} included the system in the GCVS as V923~Sco.

\citet[][hereafter F11]{fekel11} presented new high-dispersion CCD spectroscopic observations. Fifty-six radial-velocity measurements for both components led to the masses of the components; these measurements have about 0.2\% precision and result in an orbital eccentricity of
$e$ = 0.472. The continuum flux ratio of the secondary to primary component was found to be 0.754 ($\Delta$mag = 0.31) at 6430 \AA. The spectra showed that the components are slow rotators; the $v_1 \sin i$ = 5.2$\pm$1.0 km~s$^{-1}$ and $v_2 \sin i$ = 8.0$\pm$1.0 km~s$^{-1}$. A pseudosynchronous rotational velocity was estimated as 7 km~s$^{-1}$ for both components. The authors also indicated the system to be suitable for long-baseline interferometric observations. 

The search for a possible secondary eclipse in the photometry of \citet{bolton76} at the predicted spectroscopic conjunction (at phase 0.3713 from the primary eclipse) was not conclusive owing to insufficient precision and quantity of the data. The Hipparcos photometry (HIP 83491) does not cover eclipses in the system and does not show any out-of-eclipse variability (95th and 5th brightness percentiles differ by only 0.02 mag). Unfortunately, the star is brighter than the magnitude range of ASAS\footnote{All Sky Automated Survey, www.astrouw.edu.pl/}, $V$ = 8-14. Searches in other all-sky databases yielded no useful photometry.

With very precise minimum component masses, the projected size of the orbit from spectroscopy, and the possibility of resolving the visual orbit with long-baseline interferometry the system promise not only precise absolute parameters of the components but also a model-independent determination of the distance.

The present paper is organized as follows: Section~\ref{sec_observations} describes high-precision satellite photometry and VLTI interferometry of the system, Section~\ref{sec_preliminar} gives a preliminary analysis of individual datasets, and Section~\ref{sec_simultan} describes the simultaneous analysis of all datasets. The absolute parameters of the components 
and their evolutionary stage are discussed in Sections~\ref{sec_absolute} and \ref{sec_evolution}, respectively.

\section{New observations \label{sec_observations} and data reduction}

Having very precise minimum masses for the components from F11, $M_1 \sin^3 i$ = 1.4708(31) M$_\odot$ and $M_2 \sin^3 i$ = 1.4178(23) M$_\odot$, but only limited photometry \citep{bolton76} the system was subject to new photometric and interferometric observations.

\subsection{MOST photometry \label{sec_MOSTdata}}

The MOST (Microvariability and Oscillations of STars) microsatellite houses a 15-cm telescope, which feeds a CCD photometer through a single, broadband optical filter (350–-700 nm). The initial post-launch performance was described by \citet{matthews04}. Although the original mission goals were asteroseismology of bright ($V < 6$) solar-type stars, Wolf-Rayet and magnetic pulsating stars, MOST obtained photometry of all kinds of variable stars, transiting exoplanets but also of a few dozens of eclipsing binary stars \citep[see][]{pribulla10}.

Unfortunately, V923~Sco is outside the so-called continuous viewing zone; thus it was alternated with other objects to charge MOST batteries using built-in solar panels \citep[see][]{walker03}. Hence the photometry is noncontinuous and the data segments repeat with the orbital period of the satellite (101.4 minutes). The new observations were focused on the eclipses in the system using the spectroscopic prediction of F11.

The MOST frames were reduced in the following steps: (1) clipping two-sigma outliers in counts and telescope pointing position, which was repeated twice; (2) removing the power at all frequencies less than 3 c/d that have S/N $>$ 10; (3) filtering of sky background 
and inter-pixel variations using polynomials; (4) filtering of orbital modulation of scattered light by removing patterns averaged across 35 bins phased at the MOST satellite orbital period; (5) clipping three-sigma outliers in data averaged in 50-point bins; and (6) restoring power removed in step 2.

In 2012 the photometry was obtained from May 12 to May 18 (the secondary eclipse) and from May 31 to June 12 (the primary). Unfortunately, the 2012 data suffer from a saturation of the CCD at maximum light. Only part of the primary minimum is below the saturation level and could be used. The observations were therefore repeated in 2014 on May 1 and 2 (the primary eclipse) and 
July 23 and 24 (the secondary). The out-of-eclipse light curve shows about 1.8 mmag scatter (as determined from the best fit). The photometry covers both minima. Because of the eccentric orbit the primary minimum is about 0.38 mag deep and partial, while the secondary is a grazing eclipse with only $\Delta$mag = 0.013. The presence of the secondary eclipse is, however, crucial for reliable element determination. 

\subsection{Long-baseline interferometry \label{sec_VLTIdata}}

Using the projected size of the orbit from F11, $a \sin i$ = 0.2971(2) au, and the Hipparcos parallax \mbox{$\pi$ = 15.46(40) mas} \citep{leeuwen07}\footnote{Original Hipparcos astrometry solution resulted in \mbox{$\pi$ = 15.61(80) mas}}, the expected apparent size of the orbit is $a \sin i$ = 4.59 mas. The angular sizes of the components assuming $R_1$ = 2.0 R$_\odot$ and $R_2$ = 1.9 R$_\odot$ (F11) are expected to be only 0.29 and 0.27 mas. 

\begin{table*}[t]
\caption{Relative astrometric position, $\Delta X$ and $\Delta Y$, of the secondary component of V923~Sco determined from the VLTI observations. 
The orbital phases ($\phi$) correspond to the ephemeris HJD 2\,456\,779.83651 + 34.838646 $\times E$ defined by the primary minimum from MOST photometry in 2014 and the orbital period from spectroscopy (F11). The table also gives the flux ratios $F_2/F_1$ of the components (the error of the last digit is given in parenthesis) and the corresponding photometric band(s). The flux ratio is the average for the photometric bands used. The uncertainty of the position is expressed by the error ellipse with major and minor axes $a, b$ and the position angle measured from north through east.\label{tab_journal}}
\begin{center}
\begin{tabular}{ccllccrllll}
\hline \hline
HJD          & $\phi$ & ~~~$\Delta X$ & ~~~$\Delta Y$ & $a$  & $b$  &   P.A.   & $F_2/F_1$ & Bands & Stations & Combiner \\ 
             &        &  ~~[mas]      & ~~[mas]       &  ~~[mas]    & ~~[mas]  & [deg]  &       &          &          \\
\hline
2456388.877 & 0.758 & $-$4.251  &   +1.850 &   0.2633 & 0.1250 & 113.5 & 0.837(11)  & $H+K$ & A1-G1-J3    & AMBER   \\ 
2456507.582 & 0.165 &   +2.635  & $-$1.339 &   0.1936 & 0.1250 & 116.9 & 0.829(4)   & $H+K$ & A1-G1-J3    & AMBER   \\ 
2456540.584 & 0.112 &   +3.206  & $-$1.548 &   0.1132 & 0.0125 & 115.8 & 0.9(3)     & $H$   & A1-G1-K0-J3 & PIONIER \\ 
2456727.837 & 0.487 & $-$3.187  &   +1.114 &   0.4082 & 0.1250 & 109.3 & 0.789(7)   & $H+K$ & A1-G1-J3    & AMBER   \\ 
2456728.852 & 0.516 & $-$2.849  &   +0.894 &   0.1947 & 0.1250 & 107.4 & 0.899(20)  & $H+K$ & A1-G1-J3    & AMBER   \\ 
2456740.864 & 0.861 & $-$3.550  &   +1.587 &   0.2311 & 0.1250 & 114.1 & 0.840(11)  & $H+K$ & A1-G1-J3    & AMBER   \\ 
2456817.776 & 0.069 &   +2.618  & $-$1.193 &   0.1905 & 0.1250 & 114.5 & 0.811(14)  & $H+K$ & A1-G1-J3    & AMBER   \\ 
2457932.627 & 0.089 &   +2.853  & $-$1.306 &   0.0300 & 0.0240 &  24.6 & 0.820(10)  & $K$   & A0-G1-J2-J3 & GRAVITY \\ 
2457936.587 & 0.203 &   +2.572  & $-$1.327 &   0.0128 & 0.0125 & 124.1 & 0.814(8)   & $K$   & A0-G1-J2-K0 & GRAVITY \\ 
\hline
\end{tabular}
\end{center}
\end{table*}

We observed V923~Sco over 16 epochs spread between April 2013 and June 2014, using the Very Large Telescope Interferometer \citep{berger10,merand14} equipped with the three-telescope beam combiner AMBER \citep{petrov07}, or the four-telescope beam combiner PIONIER \citep{lebouq11}. The 1.8 meter auxiliary telescopes were positioned on the A1-G1-J3 and D0-H0-G1 baseline triplets for the AMBER observations, and the A1-G1-K0-J3 quadruplet for the single PIONIER observing
epoch. These configurations provide ground baselines between 56 and 140 meters, which are suitable for the resolution of the angular separation of the two components of V923~Sco. We also observed V923 Sco using the GRAVITY instrument \citep{gravity17} of the VLTI in June and July 2017. The operational wavelength band of GRAVITY is the K band.
The science combiner includes a dispersive element that provides wavelength-dependent measurements of the interferometric quantities (e.g., fringe visibilities and phases) at spectral resolutions of $R$ = 20, 500, and 4500. For our observations of V923~Sco, GRAVITY was used in medium spectral resolution ($R = 500$) and in single field mode; that is, its two beam combiners were recording fringes on V923~Sco.

The system was well resolved during nine epochs. The disks of the components were unresolved. Each observation consisted of three pointings: calibrator, V923 Sco, and calibrator. Two stars were used as calibrators: HD153368 \citep[$\theta_H = 1.005\pm$0.014 mas, $\theta_K = 1.009\pm$0.014 mas\footnote{$\theta_H$, $\theta_K$ - uniform disk angular diameter in the H and K passbands};][~the AMBER and the GRAVITY observations]{cal05}, and HD159941 \citep[$\theta_H = 1.081\pm$0.015 mas, $\theta_K = 1.089\pm$0.015 mas;][~the AMBER and the PIONIER observations]{cal05}. Each pointing took 20 minutes, leading to a total of 1 hour per observation. The list of observations with the corresponding configurations is presented in Table~\ref{tab_journal}. 

The AMBER raw data have been reduced using {\tt amdlib v3}\footnote{http://www.jmmc.fr/data\_processing\_amber.htm}. The PIONIER raw data were reduced using {\tt pndrs}\footnote{http://www.jmmc.fr/data\_processing\_pionier.htm}. The GRAVITY data were reduced using the standard GRAVITY Data Reduction System version 1.0 \citep{lapeyr14}. The binary separations were computed from the reduced files for each epoch using CANDID \citep{candid}. Because the closure phase was always achieved (three baselines for the AMBER and six for the PIONIER instrument) giving the measure and direction of the object asymmetry, the secondary position is known without the $\pm$180 degrees ambiguity affecting the case of two baselines or speckle interferometry. 

\begin{figure*}[ht]
\centering
\includegraphics[width=\textwidth]{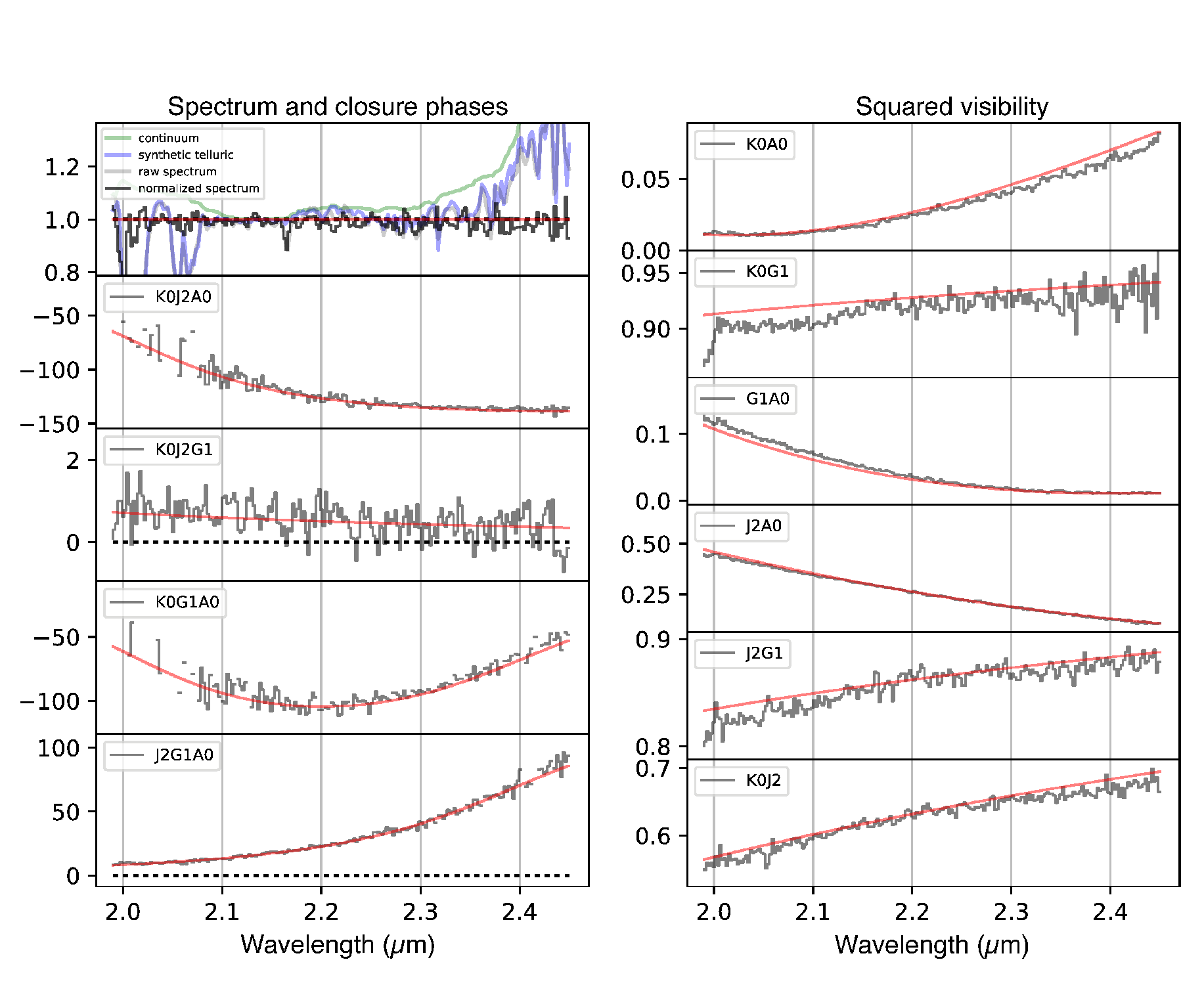}
\caption{Closure phases (left column) and squared visibilities (right column) of V923 Sco measured with GRAVITY on 2 July 2017 (HJD=2457936.587). The best fit CANDID binary star model is represented with red curves in each panel. The recorded photometric spectrum is also shown in the upper left panel before and after the correction of the telluric absorption lines. \label{gravity-fit}.}
\end{figure*}

As an illustration, Fig.~\ref{gravity-fit} shows the result of the CANDID adjustment of a binary star model (red curves) to the interferometric observables produced by the GRAVITY instrument for the observation recorded on 2 July  2017.

The errors of the relative positions derived from interferometric observations strongly depend on the direction on the sky. This results from insufficient and nonuniform coverage of the aperture and causes the synthesized PSF to be elongated at a certain angle. Hence errors of all positional measurements were represented by ellipses. Their major axis and orientation were determined by bootstrapping of the data. For all epochs 1000 bootstrapping experiments were performed. 

An additional source of position uncertainty of the interferometric observations is the precision of the wavelength calibration. The extra errors were estimated at 5\%, 2.5\%, and 0.1\% for the AMBER, PIONIER, and GRAVITY beam combiners, respectively. The wavelength uncertainty was taken into account in the error ellipses.

The relative positions of the secondary component and the flux ratio ($F_2/F_1$) are listed in Table~\ref{tab_journal}. The average flux ratio in the $H + K$ photometric band, $F_2/F_1$ = 0.82$\pm$0.02, is larger than the spectroscopic flux ratio. This means that the secondary component is redder (=cooler) than the primary.

\subsection{Observed colors and surface temperatures \label{sec_colours}}

According to the 2MASS catalog \citep{2mass}, ${J = 5.160(29)}$, ${H = 4.981(24),}$ and ${K_s = 4.895(20)}$ for V923~Sco. The observation was performed just after the primary eclipse at phase 0.00970(5) (ephemeris from Table~\ref{tab_photo}). Using the transformation of \citet{kbb88} $K_s$ transforms to $K =  4.934(20)$.

\citet{feinstein74}, studying the field of NGC~6281, listed the visual magnitudes and colors of V923~Sco as $V$ = 5.90, $B-V$ = 0.40, and $U-B = -0.02$ (three photoelectric observations)\footnote{Times of observations not given but are consistent with out-of-eclipse magnitudes of \citet{bolton76}} and mentioned that it is a nonmember object. 
The authors determined the distance to the cluster at \mbox{$d = 560\pm$30 pc}, and interstellar reddening of $E_{(B-V)}$ = 0.15$\pm$0.02. Assuming a uniform distribution of the interstellar material toward NGC~6281, and the Hipparcos distance ($64.7^{+1.7}_{-1.6}$ pc), leads to $E_{(B-V)}$ = 0.0173(25) for V923~Sco. The error is dominated by the reddening uncertainty toward NGC~6281.

If we adopt $R_K$ = $A_K/E_{B-V}$ = 0.346 and \mbox{$R_V$ = $A_V/E_{B-V}$ = 3.09} \citep[see][]{ebv85}, and assume uncertainty of the observed $V$ magnitude as 0.01 mag, then the dereddened apparent brightness of the system is \mbox{$V$ = 5.846(12)} and $K$ = 4.928(20) and the corresponding color index $(V-K)_0$ = 0.918(23). 

The authors of F11 found that the components essentially have solar iron abundances. Their determination of masses and radii of the components corresponds to $\log g_1$ = 4.00 (cgs) and $\log g_2$ = 4.03 (cgs). Using the calibration of \citet{worthey11} for solar iron abundance and $\log g$ = 4.00 and \mbox{$(V-K)_0$ = 0.918(23)} we get $T_{\rm eff}$ = 6820$\pm$40 K. Using dereddened \mbox{$(B-V)_0$ = 0.383} leads to 6750 K. The $(B-V)_0$ error can be estimated at about 0.02, which propagates to about an 80 K error of $T_{\rm eff}$. 
The effective temperature corresponds to the combined color of the components.

\section{Separate analysis of datasets \label{sec_preliminar}}

\begin{figure*}[ht]
\centering
\includegraphics[width=\textwidth]{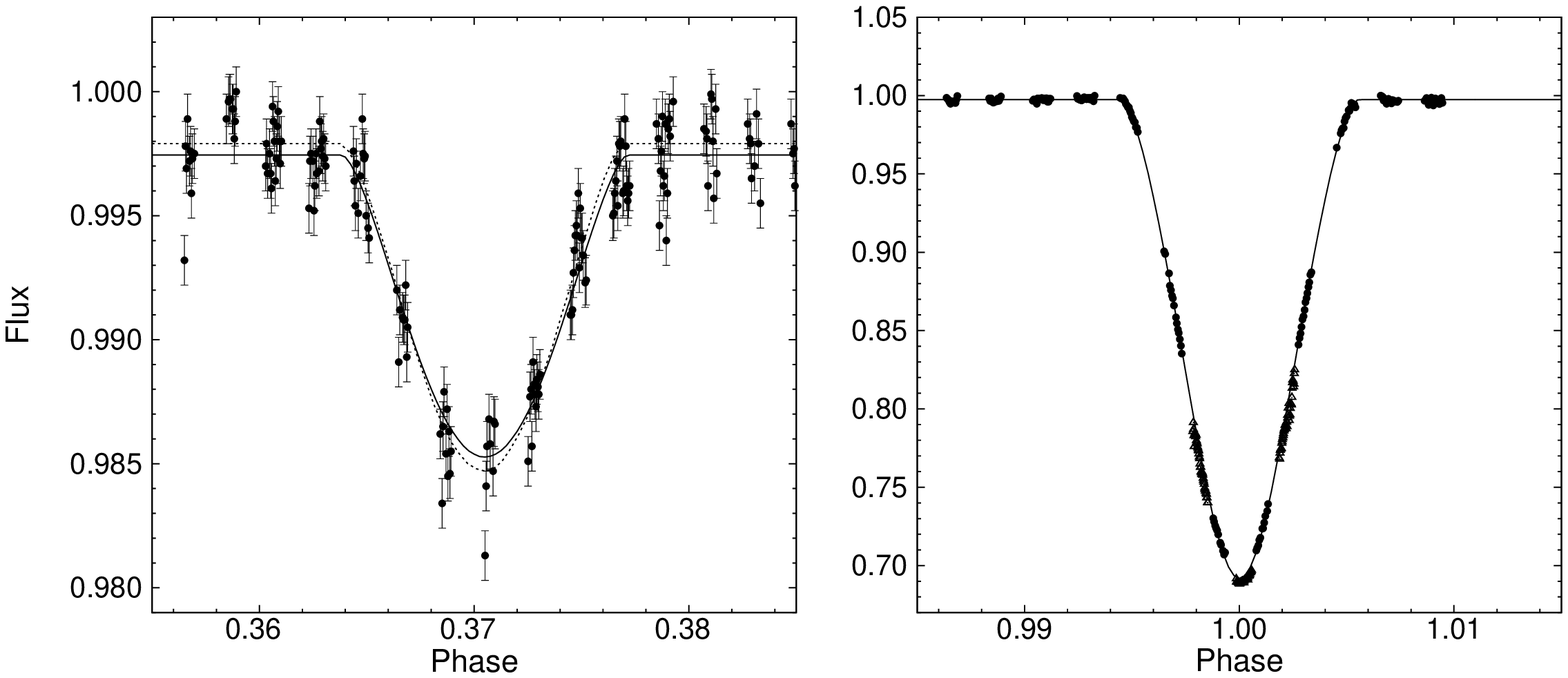}
\caption{Best fits to the secondary (left) and primary (right) minimum of V923~Sco. The 2014 data are plotted with filled circles, the nonsaturated 2012 data with open triangles. For clarity the data errors are not shown during the primary minimum (smaller than the symbols). The phases correspond to the optimum ephemeris for the primary minimum in Table~\ref{tab_photo}.
The best fit assuming two limb-darkened spheres is plotted with a solid line. The dotted line corresponds to a simultaneous modeling of the light-curve, radial-velocity and visual-orbit data (Section \ref{sec_simultan}). \label{fig_MOST}}
\end{figure*}

The published radial velocities, the MOST light curve and the interferometric visual orbit were first analyzed separately to obtain preliminary parameters and reasonable error estimates for the data independent of formal errors of individual observations. 

\subsection{Published spectroscopy \label{sec_RVs}}

In multi-dataset modeling of the system, the old photographic spectroscopy of \citet{bennet67} was not included because of the very large scatter. Hence, only new CCD spectroscopy of F11 was taken into account.

The authors do not list the errors of individual radial-velocity measurements, but comparing variances of the separate solutions for the primary and secondary component, adopt relative weights \mbox{$w_1$ = 1.0} and $w_2$ = 0.3, respectively. In our analysis these weights were transformed ($w = 1/\sigma^2$) to data uncertainties $\sigma_1$ = 1.00 km~s$^{-1}$ and $\sigma_2$ = 1.82 km~s$^{-1}$. 
The differential-correction optimization \citep[i.e., the steepest-descent method in][]{press86} of the radial velocities leads to practically the same spectroscopic elements as obtained by F11. The resulting \mbox{$\chi^2$ = 2.4} for (112-7) degrees of freedom (d.o.f.) indicates that the true data uncertainties are 6.62 times smaller, that is, $\sigma_1$ = 0.151 km~s$^{-1}$ and $\sigma_2$ = 0.275 km~s$^{-1}$. The parameter errors, checked by the bootstrap and Monte Carlo techniques, are almost identical to those listed by F11. The minimum masses, $M_1 \sin^3 i$ = 1.4708(31) M$_\odot$ and $M_2 \sin^3 i$ = 1.4178(23) M$_\odot$ listed in their Table 5, obtained from the same spectroscopic elements, however, correspond to the solar-mass parameter, $G M_\odot$, smaller by about 0.23\% than adopted by the International Astronomical Union (IAU). Using the IAU resolution B3 for solar and planetary properties \citep{iaub3} with \mbox{$G M_\odot$ = 1.3271244(10)~10$^{20}$ m$^3$ s$^{-2}$}, we get the minimum masses as 1.4674(31) M$_\odot$ and 1.4146(23) M$_\odot$. The radial velocities define the orbital period as $P$ = 34.838648(95) days.

\subsection{MOST light curve \label{sec_MOSTfit}}

Adopting radius estimates and the projected size of the major axis from F11 and assuming that we see the orbit edge on, the ratios of component radii and the instantaneous separation are about \mbox{$R_1/a(1-e) = 0.059$} and \mbox{$R_2/a(1-e) = 0.056$} at periastron, and only \mbox{$R_1/a(1+e) = 0.021$} and \mbox{$R_2/a(1+e) = 0.020$} at apastron. This means very small proximity effects (tidal deformation, mutual irradiation, and gravity darkening). 

Because of the slow rotation, the ellipsoidal deformation of the components is also small. If $R_{\rm equ}$ and $R_{\rm pol}$ are the equatorial and polar radii and $\Omega/\Omega_{\rm crit}$ is the ratio of the observed and critical angular velocity, the flattening of a rotating star can be approximated by \citep[see][]{roze09}

\begin{equation}
\frac{R_{\rm equ} - R_{\rm pol}}{R_{\rm pol}} = \frac{1}{2}
\left(\frac{\Omega}{\Omega_{\rm crit}}\right)^2
\end{equation}

with

\begin{equation}
\Omega_{\rm crit} = \sqrt{\frac{GM}{R_{\rm equ}^3}},
\end{equation}

\noindent where $G$ is the gravitational constant and $M$ is the mass of the star. For the secondary component of V923~Sco, which rotates faster of the two, $v_2 \sin i$ = 8 km~s$^{-1}$, the flattening is only 2.27~10$^{-4}$.

The reflection effect also has very low amplitude. The fraction of the reflected light can be simply estimated \citep[see][]{sen48} as $(1/4) r_j^2$, where $r_j$ is the fractional radius of the reflecting component (the ratio of its radius and semimajor axis). If both components have the same luminosity, then the amplitude of the reflection effect is $\approx (1/8) r_j^2$. In the case of V923~Sco this gives an amplitude $<0.4$ mmag, which is below the MOST photometry precision.

Thus for the preliminary parameter estimate it is sufficient to model the system by two spherical, limb-darkened stars revolving in an eccentric orbit. A simple program integrating light from the visible surface on the eclipsed component was used. 

\begin{table}[b]
\caption{Photometric elements derived from the MOST observations of V923~Sco. The last line gives $\chi^2$ and number of degrees of freedom (d.o.f.) for the mean error of each individual data point of 1.8 mmag. \label{tab_photo}}
\begin{center}
\begin{tabular}{lcc}
\hline \hline
Parameter       &               &  $\sigma$   \\
\hline 
$P$ [days]      & 34.838579     & 0.000008    \\
$T_{\rm min}$ [HJD]  & 2\,456\,779.83629 & 0.00012 \\
$i$ [deg]       & 87.687        & 0.005       \\
$r_1 + r_2$     & 0.06182       & 0.00014     \\
$r_2/r_1$       & 0.8965        & 0.0028      \\
$e \cos \omega$ &$-$0.1828      & 0.0004      \\
$e \sin \omega$ &+0.4487        & 0.0017      \\
$I_1$           & 0.22799       & 0.00013     \\
$I_2$           & 0.2139        & 0.0013      \\
$\chi^2$/d.o.f. & 488/(493$-$9)   &  --         \\
\hline
\hline
\end{tabular}
\end{center}
\end{table}

Unfortunately, fitting a model light curve to the observations of a detached eclipsing binary showing partial eclipses suffers from strong
correlations between the parameters \citep[see][]{south07}. The phase shift of the secondary minimum defines $e \cos \omega$ very well, while $e \sin \omega$ is less well defined by the ratio of minima duration \citep[see][, pages 328-329]{binnen60}. One can easily determine the sum of fractional radii of the components, $r_1 + r_2$, but not their ratio (and hence individual values, which are strongly correlated). The correlation stems from the fact that a depth of a partial eclipse depends primarily on the impact factor (product of separation of components and $\cos i$) and the sum of the radii. The inclination angle, on the other hand, is well determined.

To arrive at a meaningful solution, the spectroscopic flux ratio $F_2/F_1$ = 0.754 (at 6430 \AA) was used as an additional dataset to constrain the solutions. Its uncertainty was set (rather arbitrarily) to 0.001. The flux ratio of the components was determined using the following formula for the linear limb-darkening approximation \citep[see][]{gray08}:

\begin{equation}
\frac{F_2}{F_1} = \left( \frac{r_2}{r_1} \right)^2 \frac{I_2}{I_1} \frac{(1-u_2/3)}{(1-u_1/3)},
\end{equation}

\noindent where $r_1, r_2$ are the fractional radii of the components, $I_1, I_2$ are central intensities of the disks, and $u_1,u_2$ are the linear limb-darkening coefficients for the components. The coefficients were interpolated from dedicated tables of \citet{claret14} for the MOST satellite photometry as $u_1 = u_2$ = 0.6194 for $V_\xi$ = 2 km~s$^{-1}$, $T_{\rm eff}$ = 6700 K, \mbox{$\log g = 4.0$}, and [M/H] = 0.

In addition to the 2014 MOST data, only the nonsaturated part of the 2012 MOST light curve during the primary minimum has been used. Using the data from both years better defines the orbital period but also the shape of the primary minimum.

The free parameters were the time of the primary minimum $T_{\rm min}$, the inclination angle $i$, the sum of fractional radii $r_1 + r_2$, the ratio of fractional radii $r_2/r_1 < 1$, $e \cos \omega$, $e \sin \omega$, and the central intensities of the disks, $I_1$, $I_2$. Because no additional component was detected in the spectroscopy, zero third light, $l_3$, was assumed. All data points were given the same weight.

The resulting parameters are listed in Table~\ref{tab_photo} and corresponding fits in Fig.~\ref{fig_MOST}. The separate fractional radii are $r_1$ = 0.03260(9) and $r_2$ = 0.02922(8). The eccentricity is $e$ = 0.4845(15) and the longitude of the periastron passage $\omega$ = 112.17(9) degrees. The computed flux ratio of the components, $F_2/F_1$ = 0.75404, is close to the input value. 

\subsection{VLTI visual orbit \label{sec_VLTIfit}}

The interferometric orbit of V923~Sco is the relative orbit of the fainter component around the brighter component of the eclipsing pair. Although the orientation toward the secondary component is determined without the 180-degree ambiguity, the longitude of the ascending angle $\Omega$ cannot be determined without spectroscopic observations. Because we only had the visual orbit, we would have two equally good solutions differing in $\omega$ and $\Omega$ by 180 degrees. 

The longitude of periastron in the case of the spectroscopic orbit (112.853$\pm$0.074 deg according to F11) is related to the primary component. Thus we selected the solution with $\omega \sim 293$ degrees because the relative visual orbit is the orbit of the secondary around the primary.

Nine $\Delta X$ (toward east) and $\Delta Y$ (toward north) positions when the system was resolved with the VLTI were used. The uncertainty ellipses giving the positional uncertainty in the direction of the separation vector were used.

The parameters were adjusted using the differential-correction method. Because we only had nine separation vectors, we performed two solutions fixing some parameters from spectroscopy covering 6.5 years. In the first solution only the orbital period was fixed and not adjusted. In the second solution the orbital period, eccentricity, time of the periastron passage, and longitude of the periastron were adopted.

\begin{table}[b]
\caption{Orbital elements of the visual orbit obtained from the VLTI, $P$ - orbital period, $e$ - eccentricity, $i$ - inclination angle, $T_0$ - time of periastron passage, $\omega$ - longitude of periastron, $\Omega$ - longitude of the ascending node, and $a$ - semimajor axis. Two solutions are given: (i) for $P, e, T_0, \omega$ and
(ii) for orbital period $P$ adopted from the spectroscopy.
\label{tab_visual}}
\begin{center}
\begin{tabular}{lcl|cl}
\hline
Parameter      &               &  $\sigma$  &               & $\sigma$ \\
\hline
$P$ [days]     &  34.838646    &   --       & 34.838646     &   --   \\
$e$            &  0.47204      &   --       &  0.439        &  0.014 \\
$i$ [deg]      & 87.51         &  0.23      & 87.70         &  0.26  \\
$T_0$ [HJD]    & 2454272.1636  &   --       & 2454272.00    &  0.07  \\
$\omega$ [deg] & 292.853       &   --       & 293.5         &  1.3   \\
$\Omega$ [deg] & 114.41        &  0.27      & 114.43        &  0.29  \\
$a$ [mas]      &   4.750       &  0.013     & 4.59          &  0.05  \\
$\chi^2$/d.o.f. & 23.25/(18$-$3) &   --       & 18.06/(18$-$6)  &  --    \\
\hline
\end{tabular}
\end{center}
\end{table}

\begin{figure}[h]
\centering
\includegraphics[width=8cm,angle=270]{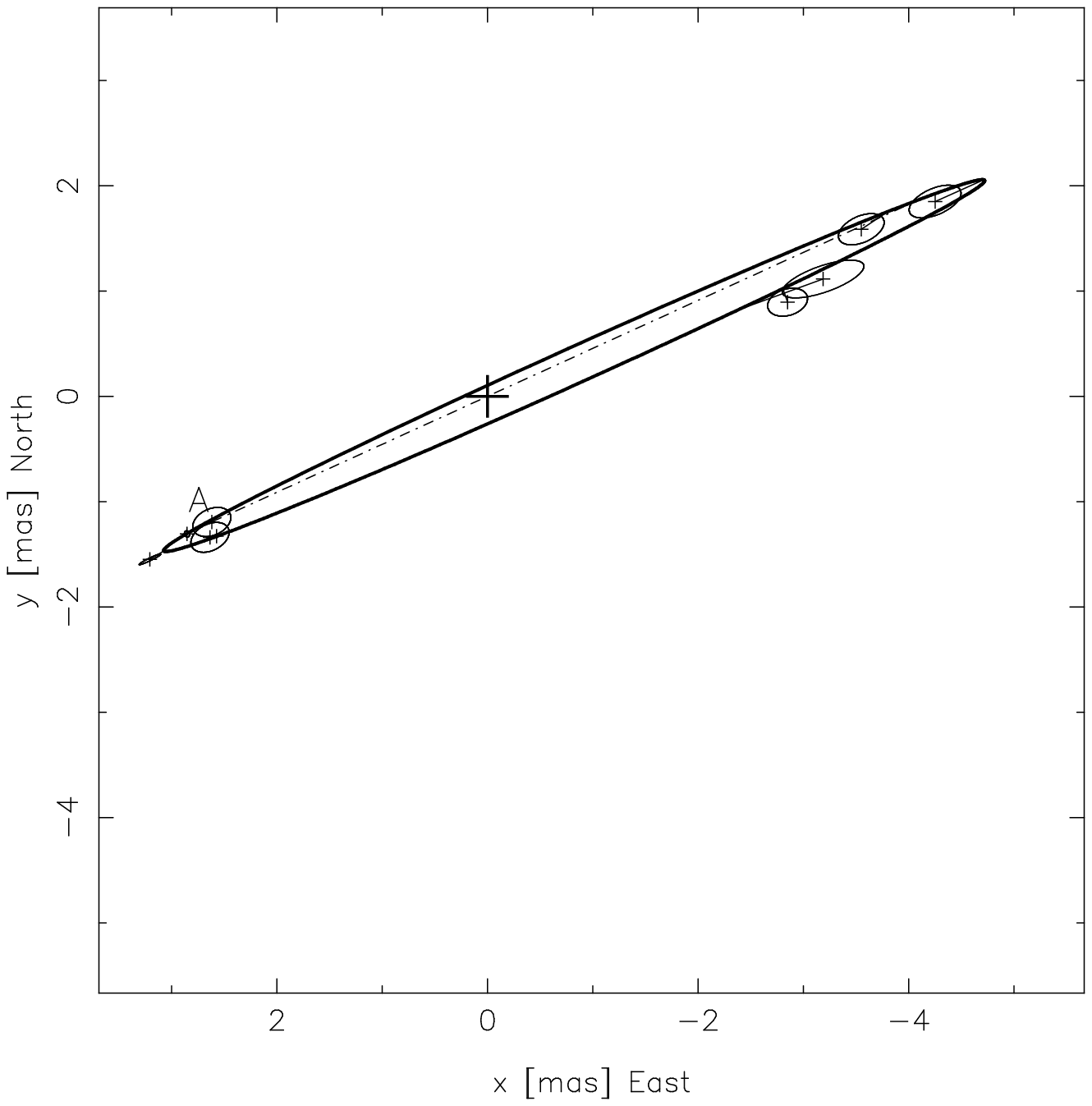}
\includegraphics[width=8cm,angle=270]{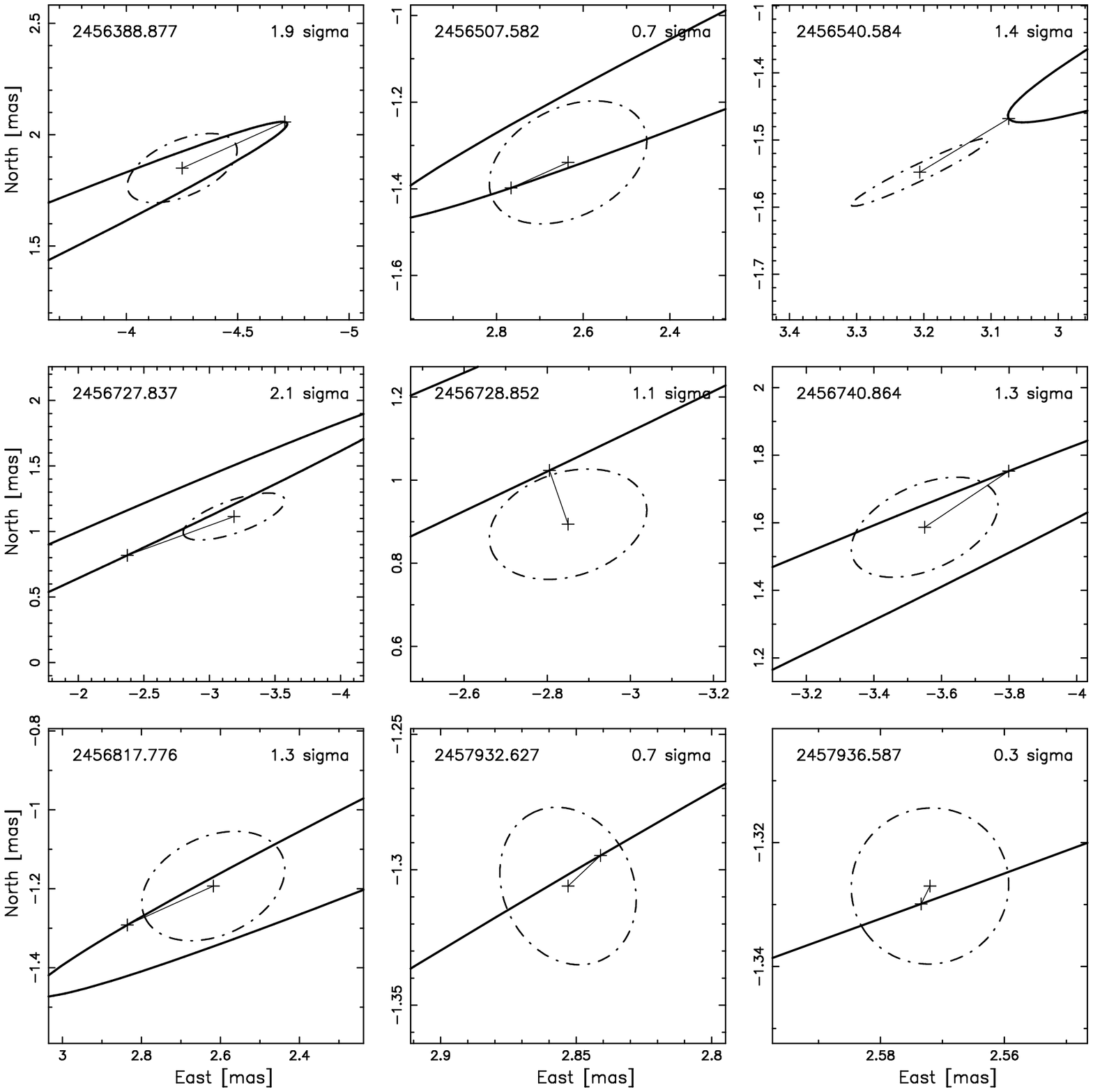}
\caption{Relative orbit of the secondary component of V923~Sco and its best fit assuming spectroscopic elements ($P, $e$, \omega, T_0$) (top). The nodal line is plotted by dash dots. Character "A" shows the position of the ascending node. Close-up view of error ellipses (dot-dashed line) at 9 epochs are shown at bottom. The orbit is prograde (counterclockwise motion on the sky). North is up and east is left on both panels \label{fig_VLTIorbit}}
\end{figure}

The corresponding fit to the relative visual orbit is shown in Fig.~\ref{fig_VLTIorbit}. The value of reduced $\chi^2_r$ = 1.505 indicates either slightly underestimated position errors or a deviating point(s). Combining the projected semimajor axis from spectroscopy, $a \sin i$ = 0.29705(19) au with apparent semimajor axis from interferometry (first solution), $a \sin i$ = 4.747(18) mas, we get the parallax of V923~Sco as 15.96(6) mas.

\section{Simultaneous modeling of the datasets \label{sec_simultan}}

The individual datasets, which were not obtained simultaneously, were light curve (HJD 2\,456\,083 and 2\,456\,863), radial velocities (2\,453\,092 - 2\,455\,465), and visual orbit (2\,456\,389 - 2\,457\,937). For a well-detached binary with $\sim$ 34.8-day period we can, however, assume that the apsidal motion is too slow to cause significant changes of orbital elements throughout the whole time range of the observations (about 13.3 years). In the case of the MOST observations, we used both 2012 (nonsaturated data in the primary minimum) and 2014 observations to define the orbital period better and to set constraints on the position of the secondary minimum.

Because the system is wide and proximity effects negligible (see Section~\ref{sec_MOSTfit}), all observations were modeled assuming that components are two limb-darkened spheres. As discussed in Section~\ref{sec_MOSTfit}, to constrain the ratio of radii, the flux ratio, $F_2/F_1$=0.754, from spectroscopy and, $F_2/F_1$ = 0.82$\pm$0.02, from interferometry were included as additional datasets. Since we did not know the uncertainty of the spectroscopic light ratio, we set this value rather arbitrarily to 0.001. The linear limb-darkening coefficient for the MOST passband was adopted from Tables 5 and 13 of \citet{claret14} after each iteration\footnote{The distance to the system was also optimized (defined primarily by the apparent size of the visual orbit and the size of the spectroscopic orbit).}. The minima last only about 0.013 (secondary) and 0.011 (primary) of an orbital period. Hence, while w used outside the minima a 1/360 step in phase, we used a phase step that was seven times finer during the minima to synthesize the light curve and photocentric radial velocity.

The multi-dataset analysis requires proper definition of the merit function to be minimized. The most obvious is to use reduced $\chi^2_r$, which for the case of combining light-curve and radial-velocity data is written as

\begin{equation}
\label{ch2a}
\chi^2_r = \frac{1}{N_{LC}+N_{RV}} \left[\sum_{i=1}^{N_{LC}} \frac{[y_i (m) - y_i (o)]^2}{\sigma_i^2} +
\sum_{i=1}^{N_{RV}} \frac{[z_i (m) - z_i (o)]^2}{\sigma_i^2} \right],
\end{equation}

\noindent where $N_{LC},N_{RV}$ are the numbers of data points for light-curve and radial-velocity data, $y$ are intensities, $z$ are radial velocities, while "m" denotes model and "o" observed. If the number of data points is markedly different between the datasets and there is a discrepancy in optimal parameters, the optimization gives higher weight to the dataset with the larger number of points. A solution to avoid this problem is to give all datasets the same weight; this can be carried out by modifying Eqn. \ref{ch2a} to

\begin{equation}
\label{ch2b}
\chi^2_r = \frac{1}{N_{LC}} \sum_{i=1}^{N_{LC}} \frac{[y_i (m) - y_i (o)]^2}{\sigma_i^2} +
\frac{1}{N_{RV}} \sum_{i=1}^{N_{RV}} \frac{[z_i (m) - z_i (o)]^2}{\sigma_i^2}.
\end{equation}


\begin{table*}
\caption{Multi-dataset modeling of V923~Sco observables. Fixed parameters are listed without error; parameters irrelevant for a given dataset(s) are skipped. The number of data points is 493 for the MOST light curve (LC), 112 for the radial-velocity (RV) curves (56 epochs), and 18 (9 epochs) for the visual orbit (VO) and two flux ratios (FR). The parameters listed are as follows: $P$ indicates the orbital period, $T_0$ the time of  
periastron passage (listed as HJD-2454272), $i$ the inclination angle, $e$ the orbital eccentricity, $T_2$ the secondary component temperature,  $r_1, r_2$ the fractional radii of the components, $K_1, K_2$ the semiamplitudes of the radial-velocity changes, $V_0$ the mass-center velocity, $\omega$ the longitude of periastron passage, $\Omega$ the 
longitude of ascending node, $\pi$ the parallax, $T_{\rm eff2}$  the effective temperature of the secondary component, and d.o.f.  the number of degrees of freedom and reduced $\chi^2_r$ for individual datasets. The modeling of the MOST light curve included a normalization factor not listed in the table but very close to unity.
\label{tab_multi}}
\begin{center}
\begin{tabular}{lcccc}
\hline \hline
Parameter        &  LC+RV        & LC+VO      & VO+RV       & LC+RV+VO     \\
\hline   
$P$              &  34.838595(6) & 34.838646  & 34.83870(7) & 34.838593(6) \\
$T_0$ [HJD]      &   0.1716(7)   & 0.1736(10) & 0.159(4)    & 0.1719(7)    \\
$i$ [deg]        &    87.724(4)  & 87.727(4)  & 87.73(8)    & 87.725(4)    \\
$e$              &   0.47243(15) & 0.47161(19)& 0.4721(4)   & 0.47242(14)  \\
$r_1$            &    0.03182(4) & 0.03180(5) & --          & 0.03182(4)   \\
$r_2$            &    0.02897(3) & 0.02897(3) & --          & 0.02897(3)   \\
$K_1$     [km/s] &     51.673(20) &  51.65    &  51.65(3)   &  51.673(20)  \\
$K_2$     [km/s] &     53.597(20) &  53.58    &  53.58(3)   &  53.597(20)  \\
$V_0$     [km/s] &  $-$15.079(8) &   --       & $-$15.082(17) & $-$15.084(8) \\
$\omega$  [deg]  &   112.964(13) & 113.058(19)& 112.83(7)   & 112.966(13)  \\
$\Omega$  [deg]  &      --       & 114.325(28)& 114.34(10)  & 114.17(26)   \\
$\pi$     [mas]  &     --        &   --       &  15.975(26) & 16.018(9)    \\
$T_{\rm eff2}$ [K] &  6562(3)    & 6565(4)    &   --        & 6565(3)      \\
d.o.f.           &     606$-$12    &  510$-$9     &  128$-$10     & 622$-$12       \\ 
$\chi^2$(LC)     &     1.009     &  1.146     &   --        & 1.005        \\
$\chi^2$(RV)     &     1.194     &  --        &  1.064      & 1.198       \\
$\chi^2$(FR)     &     0.042     &  0.042     &   --        & 0.047       \\
$\chi^2$(VO)     &      --       &  2.001     &  1.971      & 1.990       \\
\hline
\hline
\end{tabular}
\end{center}
\end{table*}

The optimization was performed by the differential correction method. The modified sum of $\chi^2_r$ (Eqn.~\ref{ch2b}) was used as the merit function. The resulting parameters for all cases are listed in Table~\ref{tab_multi}. The fit to the MOST light curve corresponding to the combined radial-velocity and light-curve solution is plotted in Fig.~\ref{fig_MOST}. 

Most parameters are consistent within one or two $\sigma$ between the solutions. Small inconsistencies are visible, for example, in the time of  periastron passage. The light and radial-velocity curves better define the orbit orientation ($\omega$) and eccentricity than the visual orbit. The position of the secondary eclipse observed by MOST perfectly defines $e \cos \omega$, while radial-velocity curves define $e \sin \omega$. On the other hand, the parallax of V923~Sco and the longitude of the ascending angle cannot be determined without the positional data. The temperature of the secondary component was determined with a very low formal error. The error estimate is affected by an unknown error of the spectroscopic flux ratio, $F_2/F_1$. Hence, it must be used with caution. 

\section{Absolute parameters and distance \label{sec_absolute}}

The distance and temperature consistency of the component can be checked using the absolute parameters of the components and the bolometric correction. Because proximity effects can be neglected and the components are very close to a spherical shape (see Section~\ref{sec_MOSTfit}) their luminosity can be satisfactorily derived from Stefan-Boltzmann's law. The absolute parameters corresponding to the combined light-curve and radial-velocity solution are listed in Table~\ref{tab_absolut}. We used the IAU resolution B3 prescriptions for the solar and planetary properties \citep{iaub3}\footnote{$R_\odot$ = 695,700 km, $T_{\rm eff \odot}$ = 5,772 K.}

\begin{table}
\caption{Absolute parameters of the components of V923~Sco from the combined light-curve and radial-velocity curve solution. We assume $T_{\rm eff1}$ = 6750 K and $T_{\rm eff2}$ = 6562$\pm$3 K. An additional temperature uncertainty resulting from the $V-K_s$ color $\sigma (T)$ = 40 K is propagated to errors of the luminosities.
\label{tab_absolut}}
\begin{center}
\begin{tabular}{lll}
\hline
Parameter           &             &  $\sigma$ \\
\hline
$R_1$ [R$_\odot$]   &  2.0246  &  0.0026  \\
$M_1$ [M$_\odot$]   &  1.4714  &  0.0014  \\
$L_1$ [L$_\odot$]   &  7.67    &  0.18    \\ 
$\log g_1$ [cgs]    &  3.9931  &  0.0012  \\
$R_2$ [R$_\odot$]   &  1.8496  &  0.0019  \\
$M_2$ [M$_\odot$]   &  1.4186  &  0.0013  \\
$L_2$ [L$_\odot$]   &  5.72    &  0.14    \\ 
$\log g_2$ [cgs]    &  4.0558  &  0.0010  \\
$a$ [R$_\odot$]     &  63.941  &  0.014   \\
\hline
\end{tabular}
\end{center}
\end{table}

Using bolometric corrections interpolated from Table~1 of \citet{popper80}, B.C. = $-0.018$ (for $T_1$ = 6750 K) and
B.C. = $-0.026$ (for $T_2$ = 6562 K) and $M_{\rm bol\odot}$ = 4.74, we get the absolute visual magnitudes as $M_{V1}$ = 2.547$\pm$0.026 and $M_{V2}$ = 2.873$\pm$0.027 and the combined absolute brightness $M_V$ = 1.945$\pm$0.037. For the observed dereddened visual brightness $V$ = 5.846(12) this gives the distance $d$ = 60.28$\pm$1.10 pc or $\pi$ = 16.59$\pm$0.30 mas. The combined solution
of radial-velocity and visual-orbit data resulted in a model-independent parallax $\pi$ = 15.975$\pm$0.026 mas or distance $d$ = 62.59$\pm$0.09 pc. To reach a model-independent distance, the components would have to be slightly cooler and thus less luminous. The model-independent distance deviates more than one sigma from the revised Hipparcos parallax of 15.46$\pm$0.40 mas or $d$ = 64.7$^{+1.7}_{-1.6}$ pc but is reasonably consistent with the original Hipparcos data reduction ($\pi$ = 15.61$\pm$0.80 mas). The discrepancy can, partially, result from the photocenter motion in this high-eccentricity system not accounted for in the Hipparcos astrometry reductions. The photocenter semiamplitude can be found as

\begin{equation}
A = \alpha \frac{q - \frac{F_2}{F_1}}{(1+q)(1+\frac{F_2}{F_1})}
,\end{equation}

\noindent where $\alpha$ is the maximum separation, $q = M_2/M_1$ is the mass ratio, and $F_2/F_1$ is the flux ratio. If $q$ = 0.964, $\alpha$ = 5.36 mas, and $F_2/F_1$ = 0.754 (assuming that the flux ratio in the Hipparcos band was the same as the spectroscopic ratio at 6430 \AA), we have $A$ = 0.33 mas. This is not a negligible oscillation compared to the Hipparcos data precision.

The distance to V923~Sco can also be checked using the surface brightness-color relations derived from long-baseline interferometry. Relations giving empirical stellar angular diameters for a zero-magnitude star can be found in \citet{boy14}. Using the polynomial coefficients from their Table 1 for $(V-K)_0$ = 0.918(23) (Section~\ref{sec_colours}) we get the zero magnitude angular radius as $\theta_{m_V = 0} = 5.96\pm$0.08 mas. Combining the relation for the angular radius of a star (in mas),

\begin{equation}
\theta = \frac{2000 R}{214.94 d}
,\end{equation}

where $R$ is the radius of a star in solar radii and $d$ is the distance in pc, and using the obvious equation $V = -5 \log (\theta/\theta_0)$ we get for the magnitude of a star
\begin{equation}
$V$ = -4.8436 - 5 \log R + 5 \log d + 5 \log \theta_{m_V = 0}.
\end{equation} 

Assuming $d$ = 62.59 pc (our combined radial-velocity and visual-orbit solution) we have $V_1$ = 6.53(3) and $V_2$ = 6.73(3) or $V$ = 5.87(3). This is consistent with the dereddened brightness of V923~Sco, $V$ = 5.846(12).

\section{Evolutionary state \label{sec_evolution}}

The evolutionary state of the components was studied by F11 using the Padova solar-abundance tracks \citep[G2000, ][]{girardi00}. Their Fig.~3 shows that the primary component of V923~Sco is overluminous for its mass. Its luminosity corresponds to about 5\% larger theoretical mass. The secondary position was consistent with theory. 

Because of the high inclination angle, $i \sim 87.7$ degrees, the true masses of the components are only 0.24\% higher than the minimum masses of F11. Although the radii of the components of V923~Sco are now known with higher precision, the luminosity is more uncertain because of the use of color-temperature calibrations, which contain about 40 K error \citep{kervel04}. The radii of the components also depend on the flux ratio. Having both the continuum flux ratio from the spectroscopy of F11 and the $H+K$ band flux ratio from new interferometry puts a strong constraint on the modeling.

\begin{figure*}[t]
\hbox{
\includegraphics[width=0.48\textwidth]{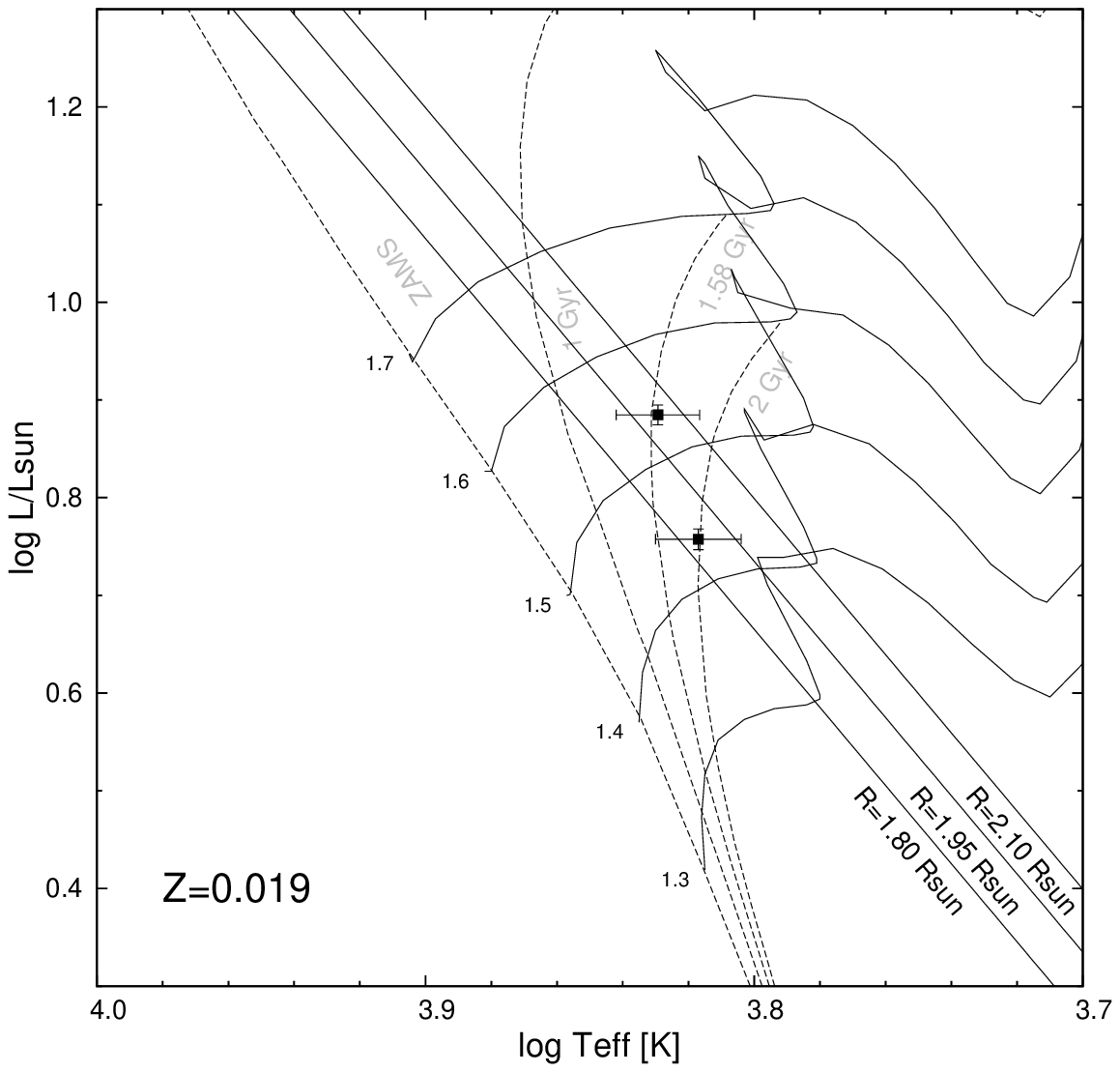}
\includegraphics[width=0.48\textwidth]{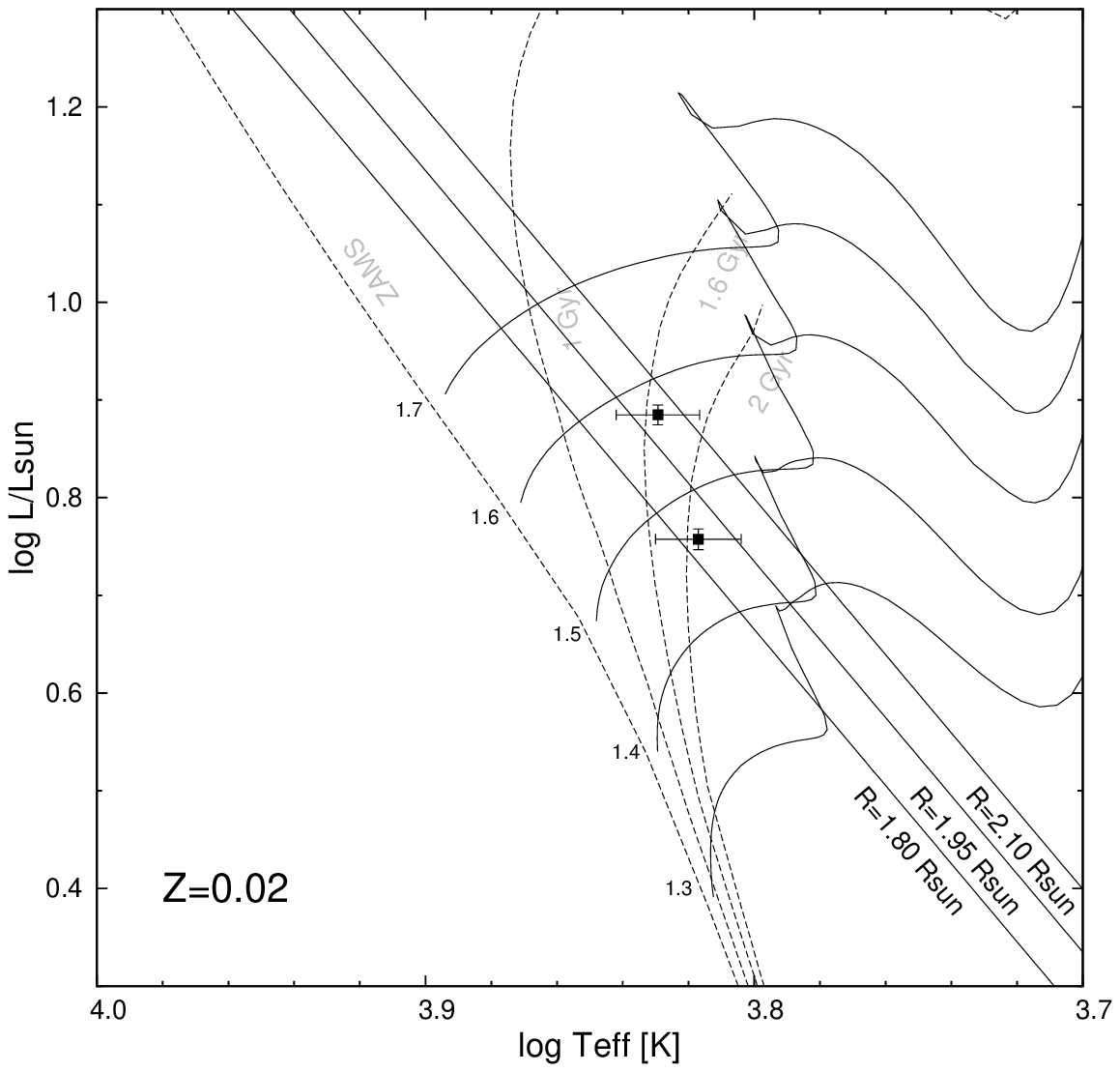}}
\caption{Hertzsprung-Russell diagram for the components of V923~Sco. The left panel shows isochrones and evolutionary tracks of \citet{girardi00} and the right panel those of \citet{YY2}. The tracks assuming convective overshooting and solar metallicity were selected in both cases. The temperature uncertainties were set to 200 K. The absolute parameters of the components correspond to the combined solution of MOST light curve and published radial-velocity data.
\label{fig_isochrones}}
\end{figure*}

While the luminosities and temperatures of the components are encumbered with the error related to color-temperature relations, their masses and radii are now known with 0.1-0.2\% errors. This is enough to perform tests of stellar evolution models \citep[see, e.g.][]{torres10}. V923~Sco was studied using (i) the G2000 and (ii) Yonsei-Yale \citep[Y$^2$, ][]{YY2} evolutionary tracks and isochrones. In both cases models corresponding to solar chemical abundance were used. Both libraries use similar physics but the Y$^2$ models differ in a number of details such as the treatment of convective core overshooting, and helium- and heavy-element diffusion. The position of the components of V923~Sco in the $T_{\rm eff} - \log g$ plane for both model libraries is shown in Fig.~\ref{fig_isochrones}. The figure shows the primary component of V923~Sco to be overluminous for its mass for both model libraries, while the parameters of the secondary component are consistent with the theory.

\begin{figure}[h]
\centering
\includegraphics[width=0.48\textwidth]{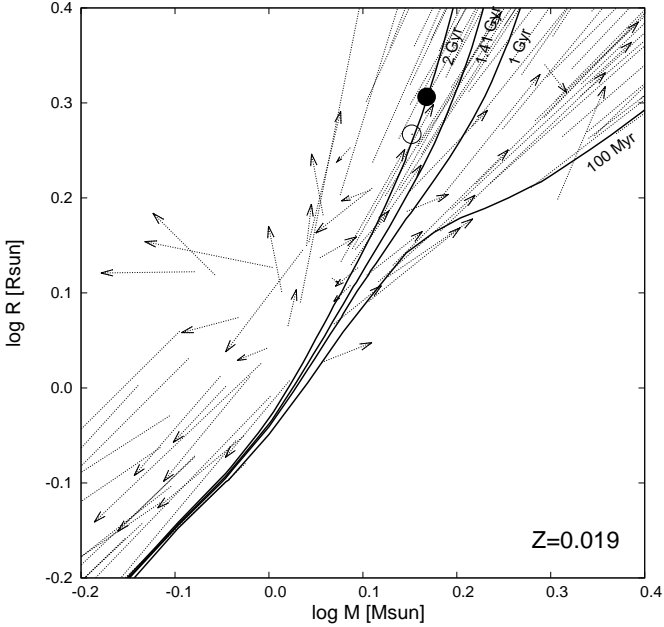}                                       
\caption{Mass-radius diagram for V923~Sco (large symbols). The measurements are compared to theoretical isochrones of \citet{girardi00} for 100 Ma to 2 Ga for solar metallicity ($Y$ = 0.70, $Z$ = 0.02). The tracks assuming convective overshooting were selected. The vectors connect the positions of secondary to primary components in other well-studied detached binaries in DEBcat of \citet{south15}.
\label{fig_debcat}}
\end{figure}

The masses and radii of the components of V923~Sco are plotted together with data of other detached eclipsing binaries listed in the online version\footnote{http://www.astro.keele.ac.uk/~jkt/debcat/} of DEBCat \citep{south15} in Fig.~\ref{fig_debcat}. The figure shows rather good agreement of the observed masses and radii for both components of V923~Sco with the theoretical isochrone for 2 Ga with about 0.1 Ga uncertainty.

\section{Discussion and conclusions}

Eclipsing binaries are crucial objects for astrophysics. Combining photometric and spectroscopic observations enables us to determine absolute parameters of stars that are necessary to test models of stellar evolution. Adding positional measurements and flux ratios (determined from spectroscopy and interferometry) gives an additional constraint and makes the results more robust.

We obtained high-precision photometry of both eclipses of V923~Sco with the MOST satellite and the spatially resolved orbit with the VLTI at nine epochs. Combining the projected size of the spectroscopic orbit, $a \sin i$, and that from the visual orbit we derive the distance to the system. Simultaneous analysis of photometric, spectroscopic, and interferometric data leads to the first reliable determination of the absolute parameters.

The main results of the combined analysis of published spectroscopic data, new satellite photometry, and VLTI interferometry of V923~Sco are as follows:

\begin{itemize}
\item First detection of the secondary eclipse of V923~Sco at phase $\phi_{II}$ = 0.370457(19), which defines the orientation of the orbit very well and sets a constraint on the photometric elements.

\item Model-independent determination of the distance to the system, $d$=62.59$\pm0.09$ pc, found by comparing $a \sin i$ from interferometric visual orbit and the projected major axis determined from spectroscopy. The error takes into account wavelength uncertainties of the interferometric combiners used.

\item Model-dependent determination of the distance based on the physical parameters and apparent brightness of the system, $d$ = 60.28$\pm$1.10 pc or $\pi$ = 16.59$\pm$0.30 mas. This indicates that the components are slightly cooler than those found by spectroscopy.
    
\item Determination of the fundamental parameters of the two components with superb accuracy (see Table~\ref{tab_absolut}).

\item The mass-radius diagram for the system indicates that the age of the system is close to 2 Ga.

\end{itemize}

The formal precision of the model-independent parallax, 26 $\mu$as, is several times worse than  can be expected for single objects in the bright-star regime of the Gaia spacecraft \citep[see][]{Gaia}. Still, V923~Sco is an important benchmark system to test the Gaia parallaxes \citep[see][]{stass16}. 

With reliable component radii the system can also be used to calibrate surface brightness-color relations \citep[see][]{graczyk16}.

Although the combined solution resulted in very accurate component parameters and orbital elements, the solution strongly depends on the spectroscopic flux ratio $F_2/F_1$ = 0.754 found by F11. The modeling, however, shows that the interferometric estimate in the near-infrared is consistent with the spectroscopic estimate. While the observed masses and radii show good accord with the 2 Ga isochrone, the primary component is overluminous for its mass. 

\begin{acknowledgements}
The authors thank the anonymous referee for valuable comments that improved the manuscript significantly. This research took advantage of the SIMBAD and VIZIER databases at the CDS, Strasbourg (France), and NASA's Astrophysics Data System Bibliographic Services. TP acknowledges support of the Fizeau exchange program (funding from the European Union’s FP7 research and innovation program under Grant Agreement 312430, OPTICON) and ESO visiting scientist program. This work has been supported by project VEGA 2/0031/18 and the Slovak Research and Development Agency under the contract No. APVV-15-0458. AFJM, JMM, and SMR are grateful for financial aid to NSERC (Canada). AFJM is also grateful for financial assistance to FQRNT (Quebec).
\end{acknowledgements}

\bibliographystyle{aa} 
\bibliography{30673} 
\end{document}